\documentclass[a4paper,12pt]{article}         
\usepackage{amsmath}
\usepackage{amssymb}
\usepackage{geometry}           
\usepackage{cite}
\usepackage{ccaption}
\usepackage{hyperref}
\usepackage{graphicx}
\usepackage[warn]{mathtext}          
\usepackage[cp1251]{inputenc}        
\usepackage[T2A]{fontenc}              
\usepackage[english]{babel}    
\usepackage{amssymb,latexsym,amsmath,amscd}

\begin{document}
\numberwithin{equation}{section}
\setcounter{footnote}{0}
\setcounter{equation}{0}
\setcounter{figure}{0}
\setcounter{table}{0}
\setcounter{section}{0}
\newcommand{\degree}{$^\circ$}

\renewcommand{\arraystretch}{1.2}
\setcounter{bottomnumber}{2}
\setcounter{totalnumber}{5}
\renewcommand{\topfraction}{0.9}
\renewcommand{\bottomfraction}{0.9}
\renewcommand{\textfraction}{0.1}

\setlength{\footskip}{1.5\baselineskip} \addtolength{\footskip}{5mm}

\renewcommand{\textfraction}{.1}
\renewcommand{\topfraction}{.9}

\newcommand{\pst}{\hspace*{1.5em}}
\newcommand{\be}{\begin{equation}}
\newcommand{\ee}{\end{equation}}
\newcommand{\bm}{\boldmath}
\newcommand{\ds}{\displaystyle}
\newcommand{\bea}{\begin{eqnarray}}
\newcommand{\eea}{\end{eqnarray}}
\newcommand{\ba}{\begin{array}}
\newcommand{\ea}{\end{array}}
\newcommand{\arcsinh}{\mathop{\rm arcsinh}\nolimits}
\newcommand{\arctanh}{\mathop{\rm arctanh}\nolimits}
\newcommand{\bc}{\begin{center}}
\newcommand{\ec}{\end{center}}
\renewcommand{\i}{\text{i}}
\newcommand{\e}{\text{e}}

\thispagestyle{plain}


\begin{center} {\Large \bf
\begin{tabular}{c}
Quadratic tomography star product algebra \\ and its classical limit
\end{tabular}
} \end{center}

\bigskip

\bigskip

\begin{center} {\bf
A. A. Strakhov$^1$ and V. I. Man'ko$^{1,2}$
}\end{center}

\medskip

\begin{center}
{\it
$^1$Moscow Institute of Physics and Technolodgy\\
Institutskii per. 9, Dolgoprudnyi, Moscow Region 141700, Russia

\smallskip

$^2$P.N. Lebedev Physical Institute, Russian Academy of Sciences\\
Leninskii Prospect 53, Moscow 119991, Russia
}
\smallskip

\end{center}

\begin{abstract}\noindent
We consider quadratic tomography in star product formalism. The contraction and the behavior of the associative algebra of quadratic tomographic symbols in $\hbar\rightarrow 0$ limit are discussed. A simple $k$-deformation example is illustrated.
\end{abstract}

\medskip

\noindent{\bf Keywords:}
star product, associative algebras, contraction, quadratic tomography.

\section{Introduction}
\pst
The conventional formulation of quantum mechanics operates with quantum states in form of density operators. First attempts to study quantum effects using purely statistical approach were made by Wigner~\cite{1}. He constructed quasiprobability distribution over the phase space and utilized it to calculate quantum corrections to classical thermodynamic averages. A bit earlier in the remarkable paper~\cite{2} Weyl introduced the correspondence of ordered operators $\hat{r}$ and $\hat{p}$ in Hilbert space to phase space functions, which is now called Weyl map.

An elaborate formulation of phase space quantization later appeared in~\cite{3}. Groenewold merged together Weyl map and Wigner function into invertible Wigner-Weyl transform between operators and phase space functions. He introduced a $\star$-product algebra of phase space functions and constructed a generalization of Poisson bracket. The classical Poisson Lie algebra was shown to be a Wigner-Inon$\ddot{\text{u}}$ contraction of the $\star$-product one. The same framework was independently discovered by Moyal~\cite{4} and the quantum Poisson bracket is now known as Moyal bracket. The evolution equation for the Wigner function was also introduced by Moyal. Further historic development of the Wigner-Weyl approach is described in~\cite{5}. Several other invertible maps from quantum operators to phase space $c$-functions were constructed~\cite{6,7,8,9}. Known functions are singular Glauber-Sudarshan quasidistribution and nonnegative Husimi quasidistribution.

The integral relation between Wigner functions and parametric probability distributions (tomograms) was found in~\cite{10}. It is used primarily to reconstruct Wigner map by measuring the state tomogram. Symplectic tomograms~\cite{11,12} were introduced to reformulate conventional density operator quantum theory into the approach of real positive marginal distribution of shifted and squeezed quadratures. Analogous tomographic procedure for spin states is based upon irreducible representations of $SU(2)$ group~\cite{13}. Symplectic tomographic map associates operators acting on Hilbert space with functions of position measured in a rotated and scaled reference frame of the phase space. These functions are called tomographic symbols. Products of operators induce $\star$-product of tomographic symbols. While the general $\star$-product associative algebras and their contractions were discussed rigorously in~\cite{14,15}, $\star$-product algebra of tomographic symbols was studied in~\cite{16}.

Recently quantum tomography was generalized to Radon transform over several nonlinear submanifolds including the quadratic ones~\cite{17}. The kernel of the $\star$-product for quadratic tomography symbols was presented in~\cite{18}.

The aim of this paper is to develop classical-to-quantum transition for the associative algebra of quadratic tomography symbols. The corresponding $\star$-product kernel will be discussed in $\hbar\rightarrow 0$ limit and the transition from noncommutative algebra to commutative will be investigated.

The article is organized as follows. In Sect. II we review the $\star$-product formalism in quantum tomography context. In Sect. III there is a short introduction to quadratic tomogram framework. In Sect. IV the kernel of quadratic tomography algebra is calculated with respect to deformation parameter. In Sect. V the first and second orders of deformed kernel are discussed. In Sect. VI a simple example of a $k$-deformation~\cite{19} of the quadratic tomography kernel presented. Finally there are some conclusions in Sect. VII.

\section{Star product by means of quantizer-dequantizer procedure}
\pst
Wigner-Weyl transform enables one to construct a nonlocal product on phase space functions corresponding to operator product. Generalizing this idea one can define a measure space $X$ and for any $x\in X$ a pair of operators $\hat{U}(x)$ and $\hat{D}(x)$ acting on Hilbert space $\mathcal{H}$ called dequantizer and quantizer respectively. One demands that 
\be
\text{Tr}[\hat{U}(x) \hat{D}(x')]=\delta(x-x').
\ee
Then to any operator $\hat{A}$ acting on $\mathcal{H}$ one can associate a function called its symbol:
\be
A(x)=\text{Tr}[\hat{A}\hat{U}(x)].
\ee
The inverse transform is performed using quantizer:
\be
\hat{A}=\int\limits_{X} A(x)\hat{D}(x)dx.
\ee
Now one can define a star product of two symbols associated with the operator product $\hat{A}\hat{B}$:
\be
A\star B(x)=\text{Tr}[\hat{A}\hat{B}\hat{U}(x)]=\int K(x',x'',x)A(x')B(x'')dx'dx'',
\ee
where the $\star$-product kernel $K(x',x'',x)$ is given by the relation
\be
K(x',x'',x)=\text{Tr}[\hat{D}(x')\hat{D}(x'')\hat{U}(x)].
\ee
Such definition of $\star$-product leads to its associativity
\be
((A\star B)\star C)(x)=(A\star (B\star C))(x),
\ee
and as a consequence an integral condition on the kernel:
\be
\int K(x_1,x_2,y)K(y,x_3,x_4)dy=\int K(x_1,y,x_4)K(x_2,x_3,y)dy.
\ee
Now let us discuss the relations between different dequantizer representations. Assume that there is a measure space $X_1$ with corresponding operators $\hat{U_1}(x_1)$, $\hat{D_1}(x_1)$, $x_1\in X_1$ and another space $X_2$ with $\hat{U_2}(x_2)$, $\hat{D_2}(x_2)$, $x_2\in X_2$. Let us associate operator $\hat{A}$ with different symbols
\be
A_1(x_1)=\text{Tr}[\hat{A}\hat{U_1}(x_1)],\;A_2(x_2)=\text{Tr}[\hat{A}\hat{U_2}(x_2)].
\ee
Then it is easy to confirm that the transition between two symbol representations is given by the following formulae:
\be
A_1(x_1)=\int\limits_{X_2}K_{12}(x_1,x_2)A_2(x_2)dx_2,
\ee
\be
A_2(x_2)=\int\limits_{X_1}K_{21}(x_2,x_1)A_1(x_1)dx_1,
\ee
where the transition functions $K_{12}$ and $K_{21}$ are:
\be
K_{12}(x_1,x_2)=\text{Tr}[\hat{U_1}(x_1)\hat{D_2}(x_2)],
\ee
\be
K_{21}(x_2,x_1)=\text{Tr}[\hat{U_2}(x_2)\hat{D_1}(x_1)].
\ee

To illustrate the presented framework let us discuss Wigner-Weyl phase space representation and symplectic tomography. We put the Planck constant $\hbar=1$, however, in order to investigate kernel contraction later let us introduce a real parameter $\mathfrak{h}\in [0,1]$ so
$$
[\hat{q},\hat{p}]=\i\mathfrak{h}.
$$
For the sake of simplicity in this paper we work only with one degree of freedom as the generalization is straightforward. 

Consider a Weyl map from the complex line to the group of unitary operators on $\mathcal{H}$:
\be
\hat{\mathcal{W}}(z)=\exp[z\hat{a}^{\dag}-z^*\hat{a}],
\ee
where 
\be
\hat{a}=\frac {\hat{q}+\i\hat{p}}{\sqrt{2\mathfrak{h}}},\;\hat{a}^{\dag}=\frac {\hat{q}-\i\hat{p}}{\sqrt{2\mathfrak{h}}},\;[\hat{a},\hat{a}^{\dag}]=1,\;
z=\frac {q+\i p}{\sqrt{2\mathfrak{h}}}.
\ee
Let us introduce Wigner-Weyl dequantizer and quantizer:
\be
\hat{U}_w(q,p)=2\hat{\mathcal{W}}(2z)\hat{\mathcal{P}},\quad \hat{D}_w(q,p)=\frac {1}{2\pi\mathfrak{h}}\hat{U}_w(q,p),
\ee
where $\hat{\mathcal{P}}$ is a parity operator. Thus the correspondence between operator $\hat{f}$ and its phase space symbol $f(q,p)$ the following:
\be
f(q,p)=2\text{Tr}[\hat{f}\hat{\mathcal{W}}(2z)\hat{\mathcal{P}}],\;\hat{f}=\frac {1} {\pi\mathfrak{h}}\int f(q,p)\hat{\mathcal{W}}(2z)\hat{\mathcal{P}}\;dqdp.
\ee
The Moyal product between phase space symbols is constructed via the Groenewold kernel:
\be
(f_1\star f_2)(q,p)=\int G(q_1,p_1,q_2,p_2,q,p)f_1(q_1,p_1)f_2(q_2,p_2)dq_1 dp_1 dq_2 dp_2,
\ee
$$
G(q_1,p_1,q_2,p_2,q_3,p_3)=\text{Tr}\Big(\hat{D}_w(q_1,p_1)\hat{D}_w(q_2,p_2)\hat{U}_w(q_3,p_3)\Big)=
$$
\be
\label{GroenewoldKernel}
\frac {1}{\pi^2\mathfrak{h}^2}\exp\left(\frac {2\i}{\mathfrak{h}}[(q_1p_2-q_2p_1)+(q_3p_1-q_1p_3)+(q_2p_3-q_3p_2)]\right).
\ee

The symplectic tomography is based upon the following pair of dequantizer and quantizer:
\be
\hat{U}_{sp}(X,\mu,\nu)=\delta (X-\mu\hat{q}-\nu\hat{p}),
\ee
\be
\hat{D}_{sp}(X,\mu,\nu)=\frac {1}{4\pi^2}\exp{\i(X-\mu\hat{q}-\nu\hat{p})},
\ee
where $X$ is a rotated and scaled quadrature with $\mu$ and $\nu$ being parameters of rotation and scaling. Let us designate $x=(X,\mu,\nu)$. The transition between tomographical symbol $w(x)$ and phase space symbol $f(q,p)$ is governed by the functions
\be
\phi(q,p,x)=\delta (X-\mu q-\nu p),
\ee
\be
\chi(q,p,x)=\frac {1}{4\pi^2}\exp{\i(X-\mu q-\nu p)},
\ee
so that
\be
w(x)=\int f(q,p)\phi(q,p,x)dqdp,\;f(q,p)=\int w(x)\chi(q,p,x)dx.
\ee
Using these relations the quantum tomogram was initially introduced as a Radon transform of a Wigner fucntion. As it was shown in~\cite{18}, the transition between the Groenewold and tomographic kernel is the following:
$$
K(x_1,x_2,x_3)=\int \chi(q_1,p_1,x_1)\chi(q_2,p_2,x_2)\phi(q_3,p_3,x_3)
$$
\be
\label{KernelTransition}
G(q_1,p_1,q_2,p_2,q_3,p_3)dq_1 dp_1 dq_2 dp_2 dq_3 dp_3.
\ee
Explicitly the kernel of symplectic tomography is the following:
$$
K(x_1,x_2,x_3)=\frac {1}{4\pi^2}\delta(\nu_3(\mu_1+\mu_2)-\mu_3(\nu_1+\nu_2))\exp\Big[\i(X_1+X_2)-\i\frac {\nu_1+\nu_2}{\nu_3}X_3+\frac {\i\mathfrak{h}}{2}(\nu_1\mu_2-\nu_2\mu_1)\Big].
$$

\section{Quadratic tomography}
\pst
Let us now move to one of the generalizations of symplectic tomogram - the quadratic one. While in the symplectic case the Radon transform is performed over all possible the affine lines, in the quadratic it is performed over all possible compact circles. Thus the parameters of the quadratic tomogram are $x=(X,\mu,\nu)$ - the radius $X$ and the phase space coordinates $\mu$ and $\nu$ of the circle center. Unlike the symplectic case the explicit form of quantizer and dequantizer is unknown, so to calculate tomographic symbols one uses Wigner-Weyl symbols and transition functions
\be
\label{Transition1}
\phi(q,p,x)=\delta(X-(q-\mu)^2-(p-\nu)^2),
\ee
\be
\label{Transition2}
\chi(q,p,x)=\frac {1}{\pi}\exp\big(\i(X-(q-\mu)^2-(p-\nu)^2)\big).
\ee
For instance the quadratic tomogram itself is 
\be
w(X,\mu,\nu)=\int W(q,p)\delta(X-(q-\mu)^2-(p-\nu)^2)dqdp,
\ee
where $W(q,p)$ is a Wigner function.
The radical difference between the symplectic and quadratic tomograms is that symplectic one is a probability distribution over $X$ while the quadratic is not as it can be negative. To prove the point let us consider a scaled harmonic oscillator $H=\frac {q^2+p^2}{2}$. The energy level Wigner functions of such oscillator are
\be
f_n=\frac {(-1)^n}{\pi\mathfrak{h}}\e^{-2H/\mathfrak{h}}L_n\Big(\frac {4H}{\mathfrak{h}}\Big),\; n=0,1,...,
\ee
where $L_n(x)$ is a Laguerre polynomial. For instance let us pick the first excited state $f_1=\frac {1}{\pi\mathfrak{h}} \e^{-2H/\mathfrak{h}}(\frac {4H}{\mathfrak{h}}-1)$. The corresponding circular tomogram for the central circles ($\mu=0$ and $\nu=0$) is the following:
\be
\omega_1(X,0,0)=\frac {1}{\pi\mathfrak{h}}\e^{-\frac {X}{\mathfrak{h}}}(\frac {2X}{\mathfrak{h}}-1).
\ee
It is easy to notice that $\omega_1(X,0,0)$ is negative for $X\in[0,\frac {\mathfrak{h}}{2}]$. 

Despite being occasionally negative quadratic tomogram is real and normalized to unity so it is a quasiprobability.

\section{Star product kernel for quadratic tomography}
\pst
In this section we will compute quadratic tomography $\star$-product kernel $\mathrm{K}$ with respect to Heisenberg Lie algebra center $\mathfrak{h}$. To obtain the kernel we will use the Groenewold kernel (\ref{GroenewoldKernel}), formula (\ref{KernelTransition}) and transition functions (\ref{Transition1}), (\ref{Transition2}) for quadratic tomography:
$$
\mathrm{K}=K(X_1,\mu_1,\nu_1,X_2,\mu_2,\nu_2,X_3,\mu_3,\nu_3)=
$$
$$
\frac {1}{\pi^6\mathfrak{h}^2}
\int\delta(X_3-(q_3-\mu_3)^2-(p_3-\nu_3)^2)
\e^{\i(X_1-(q_1-\mu_1)^2-(p_1-\nu_1)^2)}
\e^{\i(X_2-(q_2-\mu_2)^2-(p_2-\nu_2)^2)}
$$
\be
\exp\left(\frac {2\i}{\mathfrak{h}}[(q_1p_2-q_2p_1)+(q_3p_1-q_1p_3)+(q_2p_3-q_3p_2)]\right)dq_1dp_1dq_2dp_2dq_3dp_3.
\ee

The kernel $\mathrm{K}$ might be rewritten in the following manner:
$$
\mathrm{K}=\frac {\e^{\i X_1+\i X_2-\i (\mu_1^2+\nu_1^2+\mu_2^2+\nu_2^2)}}{\pi^6\mathfrak{h}^2}\int\delta(X_3-(q_3-\mu_3)^2-(p_3-\nu_3)^2)
$$
$$
\exp\left(-\i [q_1^2-2q_1(\mu_1+\frac{p_2}{\mathfrak{h}}-\frac{p_3}{\mathfrak{h}})+
(\mu_1+\frac{p_2}{\mathfrak{h}}-\frac{p_3}{\mathfrak{h}})^2-(\mu_1+\frac{p_2}{\mathfrak{h}}-\frac{p_3}{\mathfrak{h}})^2]\right)
$$
$$
\exp\left(-\i [q_2^2-2q_2(\mu_2-\frac{p_1}{\mathfrak{h}}+\frac{p_3}{\mathfrak{h}})+
(\mu_2-\frac{p_1}{\mathfrak{h}}+\frac{p_3}{\mathfrak{h}})^2-(\mu_2-\frac{p_1}{\mathfrak{h}}+\frac{p_3}{\mathfrak{h}})^2]\right)
$$
\be
\label{K1}
\e^{-\i p_1^2+2\i\nu_1p_1}
\e^{-\i p_2^2+2\i\nu_2p_2}\exp\left(\frac{2\i}{\mathfrak{h}}[q_3p_1-q_3p_2]\right)dq_1dp_1dq_2dp_2dq_3dp_3.
\ee

Let us denote $C=\frac {\e^{\i X_1+\i X_2-\i (\mu_1^2+\nu_1^2+\mu_2^2+\nu_2^2)}}{\pi^6\mathfrak{h}^2}$. Furthermore, it is now possible to evaluate a couple of Fresnel integrals using contour integration on the complex plain:
\be
\label{FresnelIntegrals}
\int\limits^{+\infty}_{-\infty}\int\limits^{+\infty}_{-\infty}
\e^{-\i(q_1-(\mu_1+\frac{p_2}{\mathfrak{h}}-\frac{p_3}{\mathfrak{h}}))^2}
\e^{-\i(q_2-(\mu_2-\frac{p_1}{\mathfrak{h}}+\frac{p_3}{\mathfrak{h}}))^2} dq_1dq_2=-\i\pi.
\ee

Substituting $C$ and (\ref{FresnelIntegrals}) in (\ref{K1}), one obtains:
$$
\mathrm{K}=-\i\pi C\int\delta(X_3-(q_3-\mu_3)^2-(p_3-\nu_3)^2)\exp\left(\i [\frac {1-\mathfrak{h}^2}{\mathfrak{h}^2} p_1^2+2p_1(-\frac {p_3}{\mathfrak{h}^2}-\frac {\mu_2}{\mathfrak{h}}+\frac {q_3}{\mathfrak{h}}+\nu_1)]\right)
$$
$$
\exp\left(\i [\frac {1-\mathfrak{h}^2}{\mathfrak{h}^2} p_2^2+2p_2(-\frac {p_3}{\mathfrak{h}^2}+\frac {\mu_1}{\mathfrak{h}}-\frac {q_3}{\mathfrak{h}}+\nu_2)]\right)\exp\left(\i (\mu_1-\frac {p_3}{\mathfrak{h}})^2+\i (\mu_2+\frac {p_3}{\mathfrak{h}})^2\right) dp_1dp_2dq_3dp_3=
$$
$$
-\i\pi C\int\delta(X_3-(q_3-\mu_3)^2-(p_3-\nu_3)^2)
$$
$$
\exp\left(\frac {\i (1-\mathfrak{h}^2)}{\mathfrak{h}^2} [(p_1-\frac {p_3+\mathfrak{h}\mu_2-\mathfrak{h} q_3-\mathfrak{h}^2\nu_1}{1-\mathfrak{h}^2})^2-\frac {(p_3+\mathfrak{h}\mu_2-\mathfrak{h} q_3-\mathfrak{h}^2\nu_1)^2}{(1-\mathfrak{h}^2)^2}] \right)
$$
$$
\exp\left(\frac {\i (1-\mathfrak{h}^2)}{\mathfrak{h}^2} [(p_2-\frac {p_3+\mathfrak{h} q_3-\mathfrak{h}\mu_1-\mathfrak{h}^2\nu_2}{1-\mathfrak{h}^2})^2-\frac {(p_3+\mathfrak{h} q_3-\mathfrak{h}\mu_1-\mathfrak{h}^2\nu_2)^2}{(1-\mathfrak{h}^2)^2}] \right)
$$
\be
\exp\left(\i (\mu_1^2+\mu_2^2)+\frac {2\i}{\mathfrak{h}^2}p_3^2-\frac {2\i}{\mathfrak{h}}(\mu_1p_3-\mu_2p_3)\right)dp_1dp_2dq_3dp_3.
\ee

Evaluating another pair of Fresnel integrals with respect to $p_1$ and $p_2$ and combining every degree of $\mathfrak{h}$ in one sentence, one will get:
$$
\mathrm{K}=(-\i\pi)\frac {\i\pi\mathfrak{h}^2}{1-\mathfrak{h}^2} C\int\delta(X_3-(q_3-\mu_3)^2-(p_3-\nu_3)^2)
\exp\Big(-\frac {\i}{\mathfrak{h}^2(1-\mathfrak{h}^2)}
[\mathfrak{h}^4 \nu_1^2+\mathfrak{h}^2 q_3^2+\mathfrak{h}^2 \mu_2^2+p_3^2+2\mathfrak{h}^3 \nu_1 q_3-
$$
$$
2\mathfrak{h}^3 \mu_2 \nu_1 - 2\mathfrak{h}^2 \nu_1 p_3 - 2\mathfrak{h}^2 \mu_2 q_3 - 2\mathfrak{h} q_3 p_3 + 2\mathfrak{h} \mu_2 p_3 +
\mathfrak{h}^4 \mu_2^2 - \mathfrak{h}^2 \mu_2^2 + 2\mathfrak{h}^3 \mu_2 p_3 - 2\mathfrak{h} \mu_2 p_3 + \mathfrak{h}^4 \nu_2^2 + \mathfrak{h}^2 q_3^2 + \mathfrak{h}^2 \mu_1^2 + p_3^2 +
$$
$$
2 \mathfrak{h}^3 \mu_1 \nu_2 - 2\mathfrak{h}^3 \nu_2 q_3 - 2\mathfrak{h}^2 \nu_2 p_3 - 2\mathfrak{h}^2 \mu_1 q_3 - 2\mathfrak{h} \mu_1 p_3 + 2\mathfrak{h} q_3 p_3 + \mathfrak{h}^4 \mu_1^2 - \mathfrak{h}^2 \mu_1^2 - 2\mathfrak{h}^3 \mu_1 p_3 + 2\mathfrak{h} \mu_1 p_3 + 2\mathfrak{h}^2 p_3^2 - 2p_3^2]\Big)
$$
$$
dq_3dp_3=\frac {\pi^2\mathfrak{h}^2C}{1-\mathfrak{h}^2}\int\delta(X_3-(q_3-\mu_3)^2-(p_3-\nu_3)^2)
\exp\Big(-\frac {\i}{1-\mathfrak{h}^2}[\mathfrak{h}^2 (\mu_1^2+\nu_1^2+\mu_2^2+\nu_2^2)+2\mathfrak{h}(\mu_1\nu_2-\mu_2\nu_1)+
$$
\be
\label{K2}
2q_3^2+2p_3^2-2\mathfrak{h}(\mu_1p_3-\mu_2p_3-\nu_1q_3+\nu_2q_3)-2(\mu_1q_3+\mu_2q_3+\nu_1p_3+\nu_2p_3)]\Big) dq_3dp_3.
\ee

Inserting back the definition of $C$ into (\ref{K2}) and simplifying the whole sentence one will obtain the following:
$$
\mathrm{K}=\frac {\e^{\i X_1+\i X_2}}{(1-\mathfrak{h}^2)\pi^4}
\int\limits_{-\infty}^{+\infty}\int\limits_{-\infty}^{+\infty}\delta(X_3-(q_3-\mu_3)^2-(p_3-\nu_3)^2)
\exp\Big(-\frac {\i}{1-\mathfrak{h}^2}\big[2\big(q_3-\frac {\mu_1-\mathfrak{h}\nu_1+\mu_2+\mathfrak{h}\nu_2}{2}\big)^2+
$$
$$
2\big(p_3-\frac {\mathfrak{h}\mu_1+\nu_1-\mathfrak{h}\mu_2+\nu_2}{2}\big)^2-
\frac {(\mu_1-\mathfrak{h}\nu_1+\mu_2+\mathfrak{h}\nu_2)^2}{2}-
\frac {(\mathfrak{h}\mu_1+\nu_1-\mathfrak{h}\mu_2+\nu_2)^2}{2}+
$$
\be
\label{K3}
(\mu_1^2+\nu_1^2+\mu_2^2+\nu_2^2)+2\mathfrak{h}(\mu_1\nu_2-\mu_2\nu_1)
\big]\Big) dq_3dp_3.
\ee

Now it is obvious, that there is a special case of $\mathfrak{h}=1$. It will be discussed a bit further, while now the calculation of the kernel $\mathrm{K}$ for $\mathfrak{h}\neq 1$ will be finished. Firstly, let us switch coordinate system from cartesian $(q_3,p_3)$ to the polar one $(r,\phi)$ in the following manner:
\be
\label{PolarSys}
q_3=r\cos\phi+\mu_3,\;p_3=r\sin\phi+\nu_3.
\ee

The implementation of the delta-functional then changes as follows:
\be
\label{DeltaInPolar}
\int\limits_{-\infty}^{+\infty}\int\limits_{-\infty}^{+\infty}\delta(X_3-(q_3-\mu_3)^2-(p_3-\nu_3)^2) dq_3dp_3=\int\limits_{0}^{2\pi}\frac {d\phi}{2\pi}
\int\limits_{0}^{+\infty}\delta(r-\sqrt{X_3})dr.
\ee

Substituting (\ref{PolarSys}) and (\ref{DeltaInPolar}) in (\ref{K3}), one obtains:
$$
\mathrm{K}=\frac {\e^{\i X_1+\i X_2}}{(1-\mathfrak{h}^2)\pi^4}\int\limits_{0}^{2\pi}\frac {d\phi}{2\pi}\exp\Big(-\frac {\i}{1-\mathfrak{h}^2}
[2X_3+(\mu_1^2+\nu_1^2+\mu_2^2+\nu_2^2+2\mu_3^2+2\nu_3^2)-2(\mu_1-\mathfrak{h}\nu_1+\mu_2+\mathfrak{h}\nu_2)\mu_3-
$$
$$
2(\mathfrak{h}\mu_1+\nu_1-\mathfrak{h}\mu_2+\nu_2)\nu_3-2\sqrt{X_3}(\mu_1-\mathfrak{h}\nu_1+\mu_2+\mathfrak{h}\nu_2-2\mu_3)\cos\phi-
$$
\be
2\sqrt{X_3}(\mathfrak{h}\mu_1+\nu_1-\mathfrak{h}\mu_2+\nu_2-2\nu_3)\sin\phi]\Big).
\ee

It is known that zero order Bessel functions can be represented as:
$$
\int\limits_{0}^{2\pi}\e^{\i a\cos\phi+\i b\sin\phi}d\phi=2\pi I_0(\sqrt{-a^2-b^2})=2\pi J_0(\sqrt{a^2+b^2}).
$$

Then, the final result for the quadratic tomography kernel $\mathrm{K}$ is:
$$
K(X_1,\mu_1,\nu_1,X_2,\mu_2,\nu_2,X_3,\mu_3,\nu_3)=\frac {\e^{\i X_1+\i X_2}}{(1-\mathfrak{h}^2)\pi^4}\exp\Big(-\frac {\i}{1-\mathfrak{h}^2}[2X_3+(\mu_1^2+\nu_1^2+\mu_2^2+\nu_2^2+2\mu_3^2+2\nu_3^2)-
$$
$$
2(\mu_1-\mathfrak{h}\nu_1+\mu_2+\mathfrak{h}\nu_2)\mu_3-2(\mathfrak{h}\mu_1+\nu_1-\mathfrak{h}\mu_2+\nu_2)\nu_3\Big)
$$
\be
J_0\Big(\frac {2\sqrt{X_3}}{1-\mathfrak{h}^2}\sqrt{(\mu_1-\mathfrak{h}\nu_1+\mu_2+\mathfrak{h}\nu_2-2\mu_3)^2+(\mathfrak{h}\mu_1+\nu_1-\mathfrak{h}\mu_2+\nu_2-2\nu_3)^2}\Big).
\ee

Let us now return to the case $\mathfrak{h}=1$. We take the relation (\ref{K3}) and perform a limit transition $\mathfrak{h}\rightarrow 1$. Using the Dirac delta regularization
$$
\delta(x)=\lim_{\epsilon\rightarrow 0}\frac {1}{\sqrt{\pi\epsilon}}\exp\Big(-\frac {x^2}{\epsilon}\Big)
$$
the formula (\ref{K3}) in the $\mathfrak{h}\rightarrow 1$ limit will be rewritten in the following manner:
$$
\mathrm{K}_{\mathfrak{h}=1}=\frac {\e^{\i X_1+\i X_2}}{2\i\pi^3}
\int\limits_{-\infty}^{+\infty}\int\limits_{-\infty}^{+\infty}\delta(X_3-(q_3-\mu_3)^2-(p_3-\nu_3)^2)\:
\delta\big(q_3-\frac {\mu_1-\nu_1+\mu_2+\nu_2}{2}\big)
$$
$$
\delta\big(p_3-\frac {\mu_1+\nu_1-\mu_2+\nu_2}{2}\big)\:dq_3 dp_3\:\lim_{\mathfrak{h}\rightarrow 1}\exp\Big(-\frac {\i}{1-\mathfrak{h}^2}\big[(1-\mathfrak{h}^2)\frac {\mu_1^2+\mu_2^2+\nu_1^2+\nu_2^2-2\mu_1\mu_2-2\nu_1\nu_2}{2}\big]\Big)=
$$
\be
\label{SimpleKernel}
\frac {2}{\i\pi^3}\e^{\i X_1+\i X_2}\e^{-\i\frac {(\mu_1-\mu_2)^2+(\nu_1-\nu_2)^2}{2}}\delta\Big(4X_3-(\mu_1-\nu_1+\mu_2+\nu_2-2\mu_3)^2-(\mu_1+\nu_1-\mu_2+\nu_2-2\nu_3)^2\Big).
\ee
This result was first obtained in~\cite{18}.

\section{The behaviour of quadratic tomography kernel for \\
infinitesimal $\mathfrak{h}$}
\pst
Let us now examine the zero and first orders of the kernel with respect to $\mathfrak{h}$ in the neighbourhood $U(0)$ of zero. This will essentially reveal classical and quantum structure of the algebra of observables in quadratic tomography scheme. Using well known Taylor series for $\mathfrak{h}\in U(0)$
\be
\frac {1}{1-\mathfrak{h}^2}=1+o(\mathfrak{h});\;\e^{-\i\mathfrak{h}}=1-\i\mathfrak{h}+o(\mathfrak{h});\;J_0(a+\mathfrak{h})=J_0(a)+o(\mathfrak{h})
\ee
one obtains zero order kernel $\mathrm{K}|_{\mathfrak{h}=0}$:
$$
\mathrm{K}|_{\mathfrak{h}=0}=
\frac {\e^{\i X_1+\i X_2}}{\pi^4}
\exp\big(-\i[2X_3+(\mu_1^2+\nu_1^2+\mu_2^2+\nu_2^2+2\mu_3^2+2\nu_3^2)-2(\mu_1+\mu_2)\mu_3-
2(\nu_1+\nu_2)\nu_3]\big)
$$
\be
J_0\big(2\sqrt{X_3}\sqrt{(\mu_1+\mu_2-2\mu_3)^2+(\nu_1+\nu_2-2\nu_3)^2}\big).
\ee

The first order $\mathrm{K}_1$ with respect to $\mathfrak{h}\in U(0)$ will be the following:
\be
\mathrm{K}_1=\mathrm{K}|_{\mathfrak{h}=0}\cdot2\i((\mu_1-\mu_2)\nu_3-(\nu_1-\nu_2)\mu_3)\mathfrak{h}.
\ee

As it is easy to notice, $\mathrm{K}|_{\mathfrak{h}=0}$ is invariant under permutation of indices $1\leftrightarrow 2$ which means $\mathrm{K}|_{\mathfrak{h}=0}$ is the kernel of associative and commutative algebra of observables $A(X,\mu,\nu)$. At the same time $\mathrm{K}_1$ changes sign under permutation $1\leftrightarrow 2$ which means that $A(X_1,\mu_1,\nu_1)\star B(X_2,\mu_2,\nu_2)\neq A(X_2,\mu_2,\nu_2)\star B(X_1,\mu_1,\nu_1)$ in general so the associative algebra $\mathrm{K}$ is noncommutative. These observations states that noncommutativity in quadratic tomography scheme is provided by quantum Planck constant $\mathfrak{h}$ just as in Wigner-Weyl phase space formalism or regular quantum mechanics in Hilbert space.

Let us now show how the quadratic star product of two tomographic symbols using classical limit kernel $\mathrm{K}|_{\mathfrak{h}=0}$ relates to classical limit star product in Wigner-Weyl scheme. We pick a pair of arbitrary phase space observables $f_1(q_1,p_1)$ and $f_2(q_2,p_2)$ and then move to their symbols in quadratic tomography scheme:
\be
A_1(X_1,\mu_1,\nu_1)=\int\limits_{-\infty}^{+\infty}\int\limits_{-\infty}^{+\infty} f_1(q_1,p_1)\delta(X_1-(q_1-\mu_1)^2-(p_1-\nu_1)^2)dq_1dp_1;
\ee
\be
A_2(X_2,\mu_2,\nu_2)=\int\limits_{-\infty}^{+\infty}\int\limits_{-\infty}^{+\infty} f_2(q_2,p_2)\delta(X_2-(q_2-\mu_2)^2-(p_2-\nu_2)^2)dq_2dp_2.
\ee

Using $\mathrm{K}|_{\mathfrak{h}=0}$ one can investigate how $A_1\star A_2$ depends on $f_1(q_1,p_1)$ and $f_2(q_2,p_2)$ in classical limit:
$$
A_1\star A_2|_{\mathfrak{h}=0}=\int\limits_{\mathbb{R}^6} dX_1d\mu_1d\nu_1dX_2d\mu_2d\nu_2\; K(X_1,\mu_1,\nu_1,X_2,\mu_2,\nu_2,X_3,\mu_3,\nu_3)|_{\mathfrak{h}=0}
$$
$$
A_1(X_1,\mu_1,\nu_1)A_2(X_2,\mu_2,\nu_2)=\int\limits_{\mathbb{R}^{10}} dX_1d\mu_1d\nu_1dX_2d\mu_2d\nu_2dq_1dp_1dq_2dp_2\;
\frac {\e^{\i X_1+\i X_2}}{\pi^4}
\exp\big(-\i[2X_3+(\mu_1^2+\nu_1^2+
$$
$$
\mu_2^2+\nu_2^2+2\mu_3^2+2\nu_3^2)-2(\mu_1+\mu_2)\mu_3-
2(\nu_1+\nu_2)\nu_3]\big) J_0\big(2\sqrt{X_3}\sqrt{(\mu_1+\mu_2-2\mu_3)^2+(\nu_1+\nu_2-2\nu_3)^2}\big)
$$
$$
\delta(X_1-(q_1-\mu_1)^2-(p_1-\nu_1)^2)\delta(X_2-(q_2-\mu_2)^2-(p_2-\nu_2)^2)f_1(q_1,p_1)f_2(q_2,p_2)=
$$
$$
\frac {1}{2\pi^5}\int\limits_{0}^{2\pi}d\phi\int\limits_{\mathbb{R}^8} d\mu_1d\nu_1d\mu_2d\nu_2dq_1dp_1dq_2dp_2\; \exp\big(\i(q_1-\mu_1)^2+\i(p_1-\nu_1)^2+\i(q_2-\mu_2)^2+\i(p_2-\nu_2)^2-2\i X_3-
$$
$$
\i(\mu_1^2+\nu_1^2+\mu_2^2+\nu_2^2)-2\i(\mu_3^2+\nu_3^2)+2\i(\mu_1+\mu_2)\mu_3+
2\i(\nu_1+\nu_2)\nu_3\big)\e^{2\i\sqrt{X_3}(\mu_1+\mu_2-2\mu_3)\cos\phi+2\i\sqrt{X_3}(\nu_1+\nu_2-2\nu_3)\sin\phi}
$$
$$
f_1(q_1,p_1)f_2(q_2,p_2)=\frac {1}{2\pi^5}\int\limits_{0}^{2\pi}d\phi\int\limits_{\mathbb{R}^8} d\mu_1d\nu_1d\mu_2d\nu_2dq_1dp_1dq_2dp_2\;
\e^{\i(q_1^2+p_1^2+q_2^2+p_2^2)-2\i X_3-2\i(\mu_3^2+\nu_3^2)}
$$
$$
\e^{2\i\mu_1(\sqrt{X_3}\cos\phi+\mu_3-q_1)}\e^{2\i\nu_1(\sqrt{X_3}\sin\phi+\nu_3-p_1)}
\e^{2\i\mu_2(\sqrt{X_3}\cos\phi+\mu_3-q_2)}\e^{2\i\nu_2(\sqrt{X_3}\sin\phi+\nu_3-p_2)}
\e^{-4\i\mu_3\sqrt{X_3}\cos\phi-4\i\nu_3\sqrt{X_3}\sin\phi}
$$
\be
\label{ClassicalStar}
f_1(q_1,p_1)f_2(q_2,p_2).
\ee

Recalling a Fourier transform representation of delta function $\delta(x)=\frac {1}{2\pi}\int\limits_{-\infty}^{+\infty}\e^{\i kx}dx$ for (\ref{ClassicalStar}) one will obtain:
$$
A_1\star A_2|_{\mathfrak{h}=0}=\frac {1}{2\pi}\int\limits_{0}^{2\pi}d\phi\int\limits_{\mathbb{R}^4}
dq_1dp_1dq_2dp_2\; \e^{\i(q_1^2+p_1^2+q_2^2+p_2^2)-2\i X_3-2\i(\mu_3^2+\nu_3^2)}
\delta(\sqrt{X_3}\cos\phi+\mu_3-q_1)
$$
$$
\delta(\sqrt{X_3}\sin\phi+\nu_3-p_1)\delta(\sqrt{X_3}\cos\phi+\mu_3-q_2)\delta(\sqrt{X_3}\sin\phi+\nu_3-p_2)
\e^{-4\i\mu_3\sqrt{X_3}\cos\phi-4\i\nu_3\sqrt{X_3}\sin\phi}
$$
$$
f_1(q_1,p_1)f_2(q_2,p_2)=\frac {1}{2\pi}\int\limits_{0}^{2\pi}d\phi\;
\e^{4\i\mu_3\sqrt{X_3}\cos\phi+4\i\nu_3\sqrt{X_3}\sin\phi}
\e^{-4\i\mu_3\sqrt{X_3}\cos\phi-4\i\nu_3\sqrt{X_3}\sin\phi}
$$
$$
f_1(\sqrt{X_3}\cos\phi+\mu_3,\sqrt{X_3}\sin\phi+\nu_3)
f_2(\sqrt{X_3}\cos\phi+\mu_3,\sqrt{X_3}\sin\phi+\nu_3)=
$$
$$
\int\limits_{-\infty}^{+\infty}\int\limits_{-\infty}^{+\infty} f_1(u+\mu_3,v+\nu_3)
f_2(u+\mu_3,v+\nu_3)\delta(u^2+v^2-X_3)dudv.
$$

Let us now reverse $A_1\star A_2(X_3,\mu_3,\nu_3)|_{\mathfrak{h}=0}$ to Wigner-Weyl phase space observable $f_1\star f_2|_{\mathfrak{h}=0}$:
$$
f_1\star f_2|_{\mathfrak{h}=0}=\frac {1}{\pi^2}\int\limits_{\mathbb{R}^3} dX_3d\mu_3d\nu_3\;A_1\star A_2(X_3,\mu_3,\nu_3)|_{\mathfrak{h}=0}\;\e^{\i X_3-\i(q_3-\mu_3)^2-\i(p_3-\nu_3)^2}=\frac {1}{\pi^2}\int\limits_{\mathbb{R}^5} dX_3d\mu_3d\nu_3dudv
$$
$$
f_1(u+\mu_3,v+\nu_3)f_2(u+\mu_3,v+\nu_3)\delta(u^2+v^2-X_3)\e^{\i X_3-\i(q_3-\mu_3)^2-\i(p_3-\nu_3)^2}.
$$

Introducing new variables
$$u'=u+\mu_3,\;v'=v+\nu_3$$
and using delta function Fourier representation one will get:
$$
f_1\star f_2|_{\mathfrak{h}=0}=\frac {1}{\pi^2}\int\limits_{\mathbb{R}^4} d\mu_3d\nu_3du'dv'\;
f_1(u',v')f_2(u',v')\e^{\i(u'^2+v'^2)-\i(q_3^2+p_3^2)}\e^{2\i\mu_3(q_3-u')}\e^{2\i\nu_3(p_3-v')}=
$$
\be
\label{ClassicalProved}
\int\limits_{-\infty}^{+\infty}\int\limits_{-\infty}^{+\infty}du'dv'\;f_1(u',v')f_2(u',v')\delta(q_3-u')\delta(p_3-v')=f_1(q_3,p_3)f_2(q_3,p_3).
\ee

As it is known Wigner-Weyl star product reduces to simple multiplication of phase space functions in classical limit $\mathfrak{h}=0$. Thus statement (\ref{ClassicalProved}) justifies $\mathrm{K}|_{\mathfrak{h}=0}$ kernel as a valid classical limit of quadratic tomography kernel.

\section{An example of a $k$-deformation of a quadratic tomography kernel}
\pst
In the paper~\cite{19} the so called $k$-deformations of associative algebras were discussed both for finite and infinite dimensions. The purpose of these deformations is constructing new associative algebras using a simple mechanism of inserting an arbitrary function $k$ in the middle of the algebra's product. In case of star product algebras this will be in a following manner:
\be
(f_1\star f_2)(x)=(f_1\star k\star f_2)(x).
\ee
The kernel $S$ of the new algebra then will be as follows:
\be
S(x_1,x_2,x_3)=\int K(x_1,z,x_3)K(y,x_2,z)k(y)\:dydz.
\ee
In case of quadratic tomography the designations are $x_i=(X_i,\mu_i,\nu_i),y=(Y,\alpha,\beta),z=(Z,\gamma,\zeta)$.

In this section we will consider a simplest deformation function $k(Y,\alpha,\beta)=\delta(Y)\delta(\alpha)\delta(\beta)$ and construct a new associative algebra kernel from the purely quantum ($\mathfrak{h}=1$) kernel (\ref{SimpleKernel}):
$$
S(X_1,\mu_1,\nu_1,X_2,\mu_2,\nu_2,X_3,\mu_3,\nu_3)=\Big(\frac {2}{\i\pi^3}\Big)^2\int dY d\alpha d\beta dZ d\gamma d\zeta \;\delta(Y)\delta(\alpha)\delta(\beta)\e^{\i(X_1+X_2+Y+Z)}
$$
$$
\exp\Big(-\frac {\i}{2}\big[(\mu_1-\gamma)^2+(\nu_1-\zeta)^2+(\alpha-\mu_2)^2+(\beta-\nu_2)^2)\big]\Big)
$$
$$
\delta\Big(4X_3-(\mu_1-\nu_1+\gamma+\zeta-2\mu_3)^2-(\mu_1+\nu_1-\gamma+\zeta-2\nu_3)^2\Big)
$$
$$
\delta\Big(4Z-(\mu_2+\nu_2+\alpha-\beta-2\gamma)^2-(\alpha+\beta-\mu_2+\nu_2-2\zeta)^2\Big)=
$$
$$
\Big(\frac {1}{2\i\pi^3}\Big)^2 \int dZ d\gamma d\zeta \;
\e^{\i(X_1+X_2+Z)}\exp\Big(-\frac {\i}{2}\big[(\mu_1-\gamma)^2+(\nu_1-\zeta)^2+\mu_2^2+\nu_2^2)\big]\Big)
$$
$$
\delta\Big(X_3-\frac {(\mu_1-\nu_1+\gamma+\zeta-2\mu_3)^2+(\mu_1+\nu_1-\gamma+\zeta-2\nu_3)^2}{4}\Big)
$$
$$
\delta\Big(Z-(\frac {\mu_2^2+\nu_2^2}{2}+\gamma^2+\zeta^2-\mu_2\gamma+\mu_2\zeta-\nu_2\gamma-\nu_2\zeta)\Big)=
$$
$$
2\Big(\frac {1}{2\i\pi^3}\Big)^2 \int d\gamma d\zeta \;
\e^{\i(X_1+X_2)}\exp\Big(\frac {\i}{2}\big[\gamma^2+\zeta^2-2(\mu_2+\nu_2-\mu_1)\gamma+2(\mu_2-\nu_2+\nu_1)\zeta-\mu_1^2-\nu_1^2\big]\Big)
$$
$$
\delta\Big((\gamma-(\mu_3+\nu_1-\nu_3))^2+(\zeta-(\mu_3-\mu_1+\nu_3))^2-2X_3\Big)=
$$
$$
\Big(\frac {1}{2\i\pi^3}\Big)^2 \int\limits_{0}^{+\infty} \frac {rdr}{\sqrt{2X_3}}\delta(r-\sqrt{2X_3})\e^{\i(X_1+X_2)}
\int\limits_{0}^{2\pi} d\phi \exp\Big(\frac {\i}{2}\big[(r\cos\phi+\mu_3+\nu_1-\nu_3)^2+(r\sin\phi+\mu_3-\mu_1+\nu_3)^2-
$$
$$
2(\mu_2+\nu_2-\mu_1)(r\cos\phi+\mu_3+\nu_1-\nu_3)+2(\mu_2-\nu_2+\nu_1)(r\sin\phi+\mu_3-\mu_1+\nu_3)-\mu_1^2-\nu_1^2\big]\Big)=
$$
$$
-\frac {1}{2\pi^5}\e^{\i(X_1+X_2+X_3)}\exp\big[\i(\mu_3^2+\nu_3^2-\mu_1\mu_2-\nu_1\nu_2+(\mu_1\nu_2-\mu_2\nu_1)-2(\mu_1\nu_3-\mu_3\nu_1)+2(\mu_2\nu_3-\mu_3\nu_2)\big]
$$
\be
J_0\big(2\sqrt{X_3}\sqrt{(\mu_1-\mu_2-\nu_3)^2+(\mu_3+\nu_1-\nu_2)^2}\big).
\ee
It is easy to notice that a new deformed kernel $S(x_1,x_2,x_3)$ is not invariant under index permutation $1\leftrightarrow 2$ and thus noncommutative and retain quantum effects.

\section{Conclusions}
\pst
Summarizing the article, we discussed the quasiprobability nature of quadratic tomogram. The kernel of quadratic tomography star product was derived and its behavior was investigated in classical limit. Using a simple $k$-deformation a new associative algebra was obtained from quadratic tomographic one and its explicit quantumness was demonstrated.

\end{document}